\documentclass{article}
\usepackage{amssymb}
\usepackage{amsmath}
\usepackage{graphicx}

\begin{document}

\title{Perturbations of the spherically symmetric collapsar in the
relativistic theory of gravitation: axial perturbations. I}
\author{V.L.Kalashnikov \\
Photonics Institute, TU Vienna, Gusshausstr. 27/387,\\
A-1040 Vienna, Austria\\
tel: +43-158-8013-8723, fax: +43-158-8013-8799,\\
e-mail:kalashnikov@tuwien.ac.at}
\date{}
\maketitle

\begin{abstract}
Horizon solutions for the axial perturbations of the spherically
symmetric metric are analyzed in the framework of the relativistic
theory of gravitation. The gravitational perturbations can not be
absorbed by the horizon that results in the excitation of the new
type of the normal modes trapped by the Regge-Wheeler potential.
The obtained results demonstrate testable differences between the
collapsar and the black hole near-horizon physics.
\end{abstract}

\section{\protect\bigskip Introduction}

\noindent Relativistic theory of gravitation (RTG) treats the
gravitation as a symmetric second-rank tensor field on the
Minkowski background spacetime \cite{logunov1}-\cite{logunov4}.
Owing to such tensor nature, the field acts likewise a curved
effective Riemannian spacetime as it takes a place in the
Einstein's theory of gravitation (theory of general relativity,
GR). However, the bi-metric nature of the RTG requires a
nontrivial inclusion of the Minkowski metric into field equation
that violates the gauge-symmetry. As a result, the non-zero
graviton's mass appears in the theory as well as there is an
additional gauge-fixing condition eliminating the undesirable
graviton's spin-states. The non-zero graviton's mass
$m_{g}<10^{-66}\ $g does not contradict the modern astrophysical
data \cite{logunov5} though its value can be essentially below
this threshold \cite{kalashnikov1} (graviton mass bound on the
Solar System dynamics is 10$^{-54}$ g \cite{talmadge}).

The extremely small but nonzero mass of graviton excludes
singularities from the RTG due to strong repulsion at small
distances (or strong field strengths). As a result, the
\textit{black holes} (i.e. field singularities covered by the pure
space-time horizons) are not possible in the theory (see, for
example \cite{logunov6}). However, the field equations allow the
existence of the extremely dense and compact objects formed at the
final stage of the collapse (so called \textit{collapsars}).
Collapsar has a radius, which is very close to the Schwarzschild
one (see \cite{kalashnikov2}), but its further contraction is not
possible due to strong repulsion induced by the massive gravitons.
Therefore the collapsar has a surface in contrast to the black
hole and falling on this surface can entail the intensive bursts
(most probably gravitational \cite{kalashnikov2}).

The modern observational data suggesting the existence of the
black-hole-like objects do not allow distinguishing the GR's black
hole from the RTG's collapsar. Moreover, the existence of the
latter has not been proved even theoretically. First step in this
direction is to prove stability of the spherically symmetric
metric in the framework of the RTG that has been made in
\cite{kalashnikov2} for the approximated metric of Refs.
\cite{logunov2,logunov6}. The main problem here is the bi-metric
character of the equations that troubles the calculations even for
the simplest field configurations. As a result, the approximations
are required already at the initial calculational stages.

In this work we consider the axial perturbations of the
spherically symmetric metric in the framework of the RTG. The
rigid analysis is not possible even in this simple case and we
need using the near-horizon approximation. The results show the
essential difference in the near-horizon physics between the GR
and the RTG. The horizon reflects (not absorb) perturbations that
allows the gravitational modes trapped between the horizon and the
Regge-Wheeler potential barrier. This means that the scattering
properties of the collapsar differ from those of the black hole.
As a result, the basic properties of the RTG can be testable in
the feature astronomical observations of the extra-compact
objects.

\section{First-order perturbations of the spherically \protect\linebreak %
symmetric metric in the RTG}

\noindent The Logunov's equations for the massive gravitational field can be
expressed through the metric of the effective Riemannian spacetime \cite%
{logunov1}-\cite{logunov4}:

\begin{gather}
G_{\nu }^{\mu }-\frac{m^{2}}{2}\left( \delta _{\nu }^{\mu }+g^{\mu \lambda
}\gamma _{\lambda \nu }-\frac{1}{2}\delta _{\nu }^{\mu }g^{\kappa \lambda
}\gamma _{\kappa \lambda }\right) =-\frac{8\pi \varkappa }{c^{4}}T_{\nu
}^{\mu },  \label{eq1} \\
D_{\mu }\hat{g}^{\mu \nu }=0,  \notag
\end{gather}

\noindent where $g^{\mu \lambda }$\ is the metric tensor of the effective
Riemannian spacetime, $\hat{g}^{\mu \nu }=\sqrt{-g}g^{\mu \nu }$; $\gamma
^{\mu \nu }$\ is the metric tensor of the background Minkowski spacetime, $%
G_{\nu }^{\mu }$\ is the Einstein tensor, $T_{\nu }^{\mu }$\ is the matter
energy-momentum tensor, $D_{\mu }$\ is the covariant derivative in the
Minkowski spacetime; $m^{2}=\left( m_{g}c\diagup \hslash \right) ^{2}$, $%
m_{g}$ is the graviton mass; $\varkappa $ is the Newtonian
gravitational constant. The second equation excluding unwonted
spin-states for the gravitational field resembles the so-called
harmonic conditions introduced
by Fock for the isolated gravitational systems (see, for example, \cite{fock}%
).

Let us write the spherically symmetric static interval of the effective
Riemannian spacetime in spherical coordinates:

\begin{equation}
ds^{2}=U\left( r\right) dt^{2}-V\left( r\right) dr^{2}-W\left( r\right)
^{2}\left( d\theta ^{2}+\sin ^{2}\theta \ d\varphi ^{2}\right) .  \label{eq2}
\end{equation}

\noindent and normalize lengths to the Schwarzschild radius $%
r_{s}=2\varkappa M\diagup c^{2}$ ($M$ is the collapsar mass). The
approximated solution of Eqs. (\ref{eq1}) for metric (\ref{eq2})
gives an
essential deviation from the Schwarzschild solution in the vicinity of $%
r_{s} $ \cite{logunov2,logunov6}:

\begin{equation}
ds^{2}=\frac{\epsilon }{2}dt^{2}-\frac{r^{2}dr^{2}}{2\Delta (r)}-r^{2}\left(
d\theta ^{2}+\sin ^{2}\theta \ d\varphi ^{2}\right) ,  \label{eq3}
\end{equation}

\noindent where $\epsilon =\frac{1}{2}\left( \frac{2\varkappa Mm_{g}}{%
\hslash c}\right) ^{2}$, $\Delta (r)=r^{2}-r$, and we have to use $%
r_{l}=r_{s}+\chi $ instead of the Schwarzschild radius. $\chi $ is
approximately proportional to $\epsilon $: $r_{l}\approx r_{s}\left(
1+1.65\epsilon \right) $ for $\epsilon \in \left[ 10^{-51},10^{-42}\right] $%
\ \cite{kalashnikov2}. $\epsilon $ is extremely small for the stellar
objects ($\approx 10^{-44}$ for $M=10M_{\odot }$ and maximum estimation for $%
m_{g}$). However, its small but non-zero value causes an
irresistible repulsion from the Schwarzschild sphere (or rather
from the sphere of $r_{l}$
radius) due to $U\neq 0$ when $V\longrightarrow \infty $ \cite%
{logunov6,kalashnikov2}.

\begin{figure}
\centering
\includegraphics[width=12 cm]{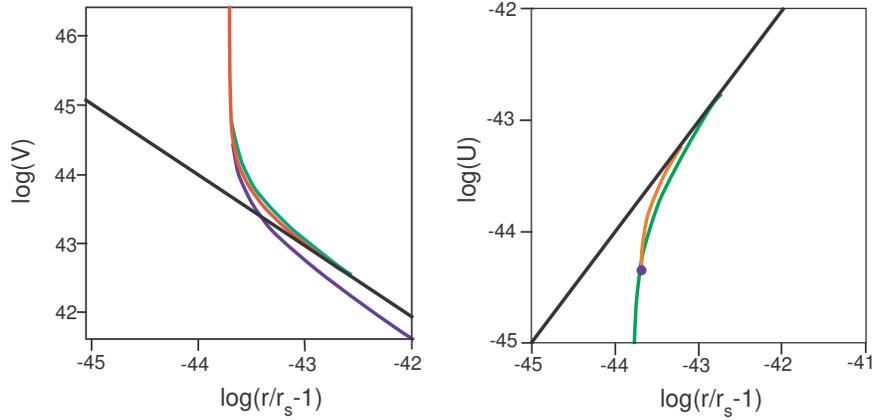}
  \caption{V (left) and U (right) functions of the interval for the Schwarzschild's (black)
  and the Logunov's (blue line and blue point) metrics. Red curves correspond to the numerically
  calculated metric. Green curves result from the approximated matched metric.}\label{pfig1}
\end{figure}

In the form of Eq. (\ref{eq3}) the spherically symmetric metric is stable \cite%
{kalashnikov2}. But this statement is not decisive for the general
metric including both Eq. (\ref{eq3}) (in the vicinity of $r_{l}$)
and Schwarzschild (out of $r_{l}$) cases. To join both solutions
we use the numerical consideration of Ref. \cite{kalashnikov2}.
Fig. \ref{pfig1} shows the dependence of $U$ and $V$ coefficients
of interval (\ref{eq2}) for the Schwarzschild (black lines) and
the Logunov (blue line and blue point, see Eq. (\ref{eq3}))
metrics. Red curve corresponds to the numerically obtained metric.
At last, the green curves result from the approximated matched
metric:

\begin{equation}
ds^{2}=\left( \frac{\epsilon }{2}+\frac{\Delta (r)}{r^{2}}\right) dt^{2}-%
\frac{r^{2}dr^{2}}{\Delta (r)}-r^{2}\left( d\theta ^{2}+\sin ^{2}\theta \
d\varphi ^{2}\right) ,  \label{eq4}
\end{equation}

\noindent which reproduces the Schwarzschild asymptotic and
simultaneously agrees with the numerical solution in the vicinity
of $r_{l}$. The main requirements for Eq. (\ref{eq4}) are i) to
replace $r_{s}$ by $r_{l}$ and ii) to consider only $r > r_{l}$.
Also, it is necessary to point at the absence of $\frac{1}{2}$ in
$V(r)$ (in comparison with Eq. (\ref{eq3})).

To analyze the axial perturbations of the metric (\ref{eq4}) we
shall use the methods of Ref. \cite{chandra}. Let us consider the
perturbed metric:

\begin{gather}
g_{\mu \nu }=  \label{eq5} \\
\begin{bmatrix}
U\left( r\right) , & 0, & 0, &
\begin{array}{c}
\zeta \ \omega \left( t,r,\theta \right) \times \\
W\left( r\right) ^{2}\sin ^{2}\theta ,%
\end{array}
\\
0, & -V\left( r\right) , & 0, &
\begin{array}{c}
\zeta \ q_{2}\left( t,r,\theta \right) \times \\
W\left( r\right) ^{2}\sin ^{2}\theta ,%
\end{array}
\\
0, & 0, & -W\left( r\right) ^{2}, &
\begin{array}{c}
\zeta \ q_{3}\left( t,r,\theta \right) \times \\
W\left( r\right) ^{2}\sin ^{2}\theta ,%
\end{array}
\\
\begin{array}{c}
\zeta \ \omega \left( t,r,\theta \right) \times \\
W\left( r\right) ^{2}\sin ^{2}\theta%
\end{array}%
, &
\begin{array}{c}
\zeta \ q_{2}\left( t,r,\theta \right) \times \\
W\left( r\right) ^{2}\sin ^{2}\theta%
\end{array}%
, &
\begin{array}{c}
\zeta \ q_{3}\left( t,r,\theta \right) \times \\
W\left( r\right) ^{2}\sin ^{2}\theta%
\end{array}%
, & -W\left( r\right) ^{2}\sin ^{2}\theta%
\end{bmatrix}%
,  \notag
\end{gather}

\noindent where $\zeta $ is the small expansion parameter defining the
perturbation amplitude; $\omega \left( t,r,\theta \right) ,~q_{2}\left(
t,r,\theta \right) $ and $q_{3}\left( t,r,\theta \right) $ are the
perturbation functions defining their form.

Let us restrict oneself to the first-order perturbations, i.e. omit all
higher than linear on $\zeta $ terms in the expansion of (\ref{eq5}).
Additionally we shall suppose the harmonic law for the time-dependence of
the perturbations:

\begin{gather}
\omega \left( t,r,\theta \right) =\widetilde{\omega }\left( r,\theta \right)
\exp \left( -i\sigma t\right) ,  \label{eq6} \\
q_{2}\left( t,r,\theta \right) =\widetilde{q}_{2}\left( r,\theta \right)
\exp \left( -i\sigma t\right) ,  \notag \\
q_{3}\left( t,r,\theta \right) =\widetilde{q}_{3}\left( r,\theta \right)
\exp \left( -i\sigma t\right) ,  \notag
\end{gather}

\noindent where $\sigma $ is some number (complex in general case).

Second equation of (\ref{eq1}), Eqs. (\ref{eq4}-\ref{eq6}) immediately lead
to the gauge-fixed condition:

\begin{gather}
\frac{6\widetilde{q}_{3}\cos \left( \theta \right) \left( r^{2}\epsilon
+2\Delta \right) }{\sin \left( \theta \right) }+\widetilde{q}_{2}\left[
2\Delta \left( 3r\epsilon +\frac{4\Delta }{r}\right) +\frac{d\Delta }{dr}%
\left( r^{2}\epsilon +4\Delta \right) \right] +  \label{eq7} \\
2\left( r^{2}\epsilon +2\Delta \right) \left[ \Delta \frac{\partial
\widetilde{q}_{2}}{\partial r}+\frac{\partial \widetilde{q}_{3}}{\partial
\theta }\right] +4ir^{4}\widetilde{\omega }\sigma =0.  \notag
\end{gather}

\noindent It is clear that for real perturbation functions and not
pure imaginary $\sigma $, Eq. (\ref{eq7}) requires
$\widetilde{\omega }=0$ (see Appendix).

Governing equations for the \textit{axial} perturbations can be
obtained from examining the $\left[ 2,4\right] $- and $\left[
3,4\right] $-components of the first Logunov's equation
(\ref{eq1}). It is convenient to use the new function:

\begin{equation}
Q\left( r,\theta \right) =\left[ \frac{\partial \widetilde{q}_{3}}{\partial r%
}-\frac{\partial \widetilde{q}_{2}}{\partial \theta }\right] \Delta \sin
^{3}\theta .  \label{eq8}
\end{equation}

\noindent Then examining the corresponding components of the Logunov's
equations results in:

\begin{eqnarray}
\frac{\left( 2\Delta +\epsilon r^{2}\right) ^{2}}{r^{2}\Delta \sin
^{3}\theta }\frac{\partial Q}{\partial \theta } &=&\widetilde{q}_{2}\left[
2\epsilon \cos ^{2}\theta ~r^{2}\Delta \left( \epsilon r+\frac{2\Delta }{r}%
\right) ^{2}+4r^{2}\Delta \sigma ^{2}\right] +  \notag \\
&&\widetilde{q}_{2}\epsilon \left[ 8r^{3}\left( 1+3\Delta -r^{3}\right)
+5-6r+2r^{4}\sigma ^{2}\right] -  \label{eq9} \\
&&\widetilde{q}_{2}\epsilon ^{2}\left[ r\left( 8r\Delta ^{2}+3\right)
+2\epsilon r^{4}\Delta \right] ,  \notag
\end{eqnarray}

\begin{eqnarray}
\frac{\Delta }{Q~r^{3}\sin ^{3}\theta }\left[ \epsilon \frac{\partial }{%
\partial r}\left( \frac{Q^{2}r^{2}}{\Delta }\right) +4Q\frac{\partial Q}{%
\partial r}\right] &=&  \label{eq10} \\
&&4\widetilde{q}_{3}\left[ \sin ^{2}\theta \left( \epsilon ^{2}r+\frac{%
2\epsilon \Delta }{r}\right) -r\sigma ^{2}\right] .  \notag
\end{eqnarray}

\noindent Eqs. (\ref{eq9},\ref{eq10}) can be reduced to a single
equation:

\begin{equation}
\frac{\partial }{\partial r}\left\{ \frac{\Delta \left[ \epsilon \frac{%
\partial }{\partial r}\left( \frac{Q^{2}r^{2}}{\Delta }\right) +4Q\frac{%
\partial Q}{\partial r}\right] }{4Q\sin ^{3}\theta ~r^{3}\Phi \left(
r,\theta \right) }\right\} +\frac{\partial }{\partial \theta }\left\{ \frac{%
2\left( 2\Delta +\epsilon r^{2}\right) ^{2}\frac{\partial Q}{\partial \theta
}}{\sin ^{3}\theta ~r^{2}\Delta \Psi \left( r,\theta \right) }\right\} =-%
\frac{Q}{\sin ^{3}\theta ~\Delta },  \label{eq11}
\end{equation}

\noindent where

\begin{eqnarray}
\Psi \left( r,\theta \right) &=&4\epsilon \cos ^{2}\theta ~r^{2}\Delta
\left( \epsilon r+\frac{2\Delta }{r}\right) ^{2}-2\epsilon ^{2}\left[
r\left( 8r\Delta ^{2}+3\right) +2\epsilon r^{4}\Delta \right] +  \notag \\
&&8r^{2}\Delta \sigma ^{2}+2\epsilon \left[ 8r^{3}\left( 1+3\Delta
-r^{3}\right) +5-6r+2r^{4}\sigma ^{2}\right] ,  \label{eq12}
\end{eqnarray}

\begin{equation}
\Phi \left( r,\theta \right) =-\sin ^{2}\theta \left( \epsilon ^{2}r+\frac{%
2\epsilon \Delta }{r}\right) +r\sigma ^{2}.  \label{eq13}
\end{equation}

\section{Horizon solution}

\noindent Eq. (\ref{eq11}) is too complicate to be solved directly and we
have to made some approximations. Let us linearize this equation on $%
\epsilon $\:

\begin{gather}
-\left( \frac{\partial }{\partial r}\frac{\Delta \frac{\partial Q}{\partial r%
}}{r^{4}\sigma ^{2}\sin ^{3}\theta }+\frac{\partial }{\partial \theta }\frac{%
\frac{\partial Q}{\partial \theta }}{r^{4}\sigma ^{2}\sin ^{3}\theta }%
\right) +  \notag \\
\frac{\epsilon }{4\sigma ^{4}\sin ^{3}\theta }\left\{ \sigma ^{2}\left[
\frac{\partial }{\partial r}\frac{Q\frac{d\Delta }{dr}}{r^{2}\Delta }-\frac{2%
}{r}\frac{\partial }{\partial r}\frac{\frac{\partial Q}{\partial r}}{r}%
\right] -8\sin ^{2}\theta \frac{\partial }{\partial r}\frac{\Delta ^{2}\frac{%
\partial Q}{\partial r}}{r^{6}}\right\} +  \notag \\
\frac{\epsilon }{r^{6}\sigma ^{4}}\left[ \frac{\Upsilon \left( r\right) }{%
4\Delta }\frac{\partial }{\partial \theta }\frac{\frac{\partial Q}{\partial
\theta }}{\sin ^{2}\theta }+2\Delta \frac{\cos ^{2}\theta }{\sin ^{2}\theta }%
\frac{\partial }{\partial \theta }\frac{\frac{\partial Q}{\partial \theta }}{%
\sin \theta }-4\Delta ^{2}\frac{\cos \theta }{\sin ^{4}\theta }\frac{%
\partial Q}{\partial \theta }\right] =  \label{eq14} \\
\frac{Q}{\sin ^{3}\theta }\left( \frac{1}{\Delta }-\frac{3\epsilon }{%
2r^{4}\sigma ^{2}}\right) ,  \notag
\end{gather}

\noindent where $\Upsilon \left( r\right) =24r^{3}\Delta -8r^{3}\left(
r^{3}-1\right) -6r+5-2r^{4}\sigma ^{2}$.

The Logunov's metrics differs from the Schwarzschild's one only in
the vicinity of the Schwarzschild's sphere where $\Delta
^{2}=O\left( \epsilon ^{2}\right) $ and $\epsilon \Delta =O\left(
\epsilon ^{2}\right) $. Hence we can neglect the corresponding
terms that results in

\begin{gather}
\frac{1}{\sigma ^{2}\sin ^{3}\theta }\frac{\partial }{\partial r}\frac{%
\Delta \frac{\partial Q}{\partial r}}{r^{4}}+\frac{1}{\sigma ^{2}r^{4}}\frac{%
\partial }{\partial \theta }\frac{\frac{\partial Q}{\partial \theta }}{\sin
^{3}\theta }-  \notag \\
\frac{\epsilon }{4\sigma ^{2}}\left[ \frac{1}{\sin ^{3}\theta }\frac{%
\partial }{\partial r}\frac{Q\frac{d\Delta }{dr}}{r^{2}\Delta }+\frac{%
5-8r^{3}\left( r^{3}-1\right) -6r-2r^{4}\sigma ^{2}}{\sigma ^{2}r^{6}\Delta }%
\frac{\partial }{\partial \theta }\frac{\frac{\partial Q}{\partial \theta }}{%
\sin ^{3}\theta }\right] =  \label{eq15} \\
\frac{Q}{\sin ^{3}\theta }\left( \frac{3\epsilon }{2r^{4}\sigma ^{2}}-\frac{1%
}{\Delta }\right) .  \notag
\end{gather}

To separate the variables in Eq. (\ref{eq15}) we consider the
limit $\Delta \longrightarrow 0$ (\textquotedblleft
horizon\textquotedblright\ solution) and take into consideration
only leading terms. This leads to:

\begin{equation}
r^{4}\frac{\partial }{\partial r}\frac{Q\frac{d\Delta }{dr}}{r^{2}\Delta }%
+\Upsilon ^{\prime }\left( r\right) \frac{\sin ^{3}\theta }{r^{2}\sigma
^{2}\Delta }\frac{\partial }{\partial \theta }\frac{\frac{\partial Q}{%
\partial \theta }}{\sin ^{3}\theta }=\frac{4r^{4}\sigma ^{2}Q}{\epsilon
\Delta },  \label{eq16}
\end{equation}

\noindent where $\Upsilon ^{\prime }\left( r\right) =\Upsilon \left(
r\right) -24r^{3}\Delta -4\ r^{2}\sigma ^{2}\Delta \diagup \epsilon $.

Substitution $Q\left( r,\theta \right) \equiv\Xi \left( r\right)
\Theta \left( \theta \right) $ gives

\begin{equation}
\frac{{d^2 \Theta }}{{dr^2 }} - 3\frac{{\cos \theta }}{{\sin
\theta }}\frac{{d\Theta }}{{dr}} + n\Theta  = 0, \label{eq17}
\end{equation}

\begin{gather}
\left[ {r^2 \left( {2r - 1} \right)} \right] \frac{{d\Xi }}{{dr}}
+\\  \left[ {\left( {\frac{{4\Delta }}{\varepsilon } + \frac{{2 +
4r\left( {r + 1} \right) - \frac{4}{r} - \frac{5}{{r^2
}}}}{{\sigma ^2 }}} \right)n - \frac{{4r^4 \sigma ^2
}}{\varepsilon } - \frac{{r^2 \left( {2r - 1} \right)^2 }}{\Delta
}} \right]\Xi  = 0,\notag  \label{eq18}
\end{gather}

Eq. (\ref{eq17}) is the well-known Gegenbauer's equation \cite{chandra} and
can be solved through the Legendre's functions. Then $n=\left( l+2\right)
\left( l-1\right) $, $l$ is an integer (angular harmonic index). $l=1$
corresponds to zero angular momentum.

Eq. (\ref{eq18}) can be integrated immediately that results in

\begin{eqnarray}
\Xi \left( r\right) &=&C\Delta \left( r\right) \left( 2r-1\right) ^{\left(
2n\left( 1/\epsilon +23/\sigma ^{2}\right) +\sigma ^{2}/\left( 2\epsilon
\right) \right) }r^{\left( -4n\left( 12/\sigma ^{2}+1/\epsilon \right)
\right) }  \label{eq19} \\
&&\times \exp \left\{ \frac{\sigma ^{2}r\left( r+1\right) }{\epsilon }+\frac{%
\left( 21\epsilon r+5\epsilon +78\epsilon r^{2}\right) n}{3r^{3}\epsilon
\sigma ^{2}}\right\} ,  \notag
\end{eqnarray}

\noindent where $C$ is the constant of integration. Eq. (\ref{eq19}) agrees
with the assumption accounting for Eq. (\ref{eq16}) because of $\Delta \frac{%
d^{2}\Xi }{dr^{2}}\longrightarrow 0$ in the considered limit.

From Eq. (\ref{eq19}) the radial perturbations have zero horizon
asymptotic, i.e. the collapsar \textquotedblleft
surface\textquotedblright\ behaves like a perfectly rigid body.
Physically this means that the collapsar's radius, mass and form
are not affected by such perturbations, i.e. such perturbations
can not be absorbed by the collapsar. It results in the crucial
difference between the RTG collapsar and the GR black hole (see
below).

For $\Delta \longrightarrow 0$ leading term of horizon solution is
(decay of the perturbation amplitude as a result of $\Delta
\longrightarrow 0$ can be obtained also in the
$\epsilon/\Delta$-order, see Appendix) :

\begin{equation}
\Xi _{L}\left( r\right) =C_{L}\left( r-1\right) ,  \label{eq20}
\end{equation}

\noindent Eq. (\ref{eq20}) has to be in accord with the Schwarzschild
horizon solution \cite{wheeler,sanchez}:

\begin{equation}
\Xi _{S}\left( r\right) =C_{S}\left( r-1\right) ^{-i\sigma }+c.c.
\label{eq21}
\end{equation}

\noindent \qquad Equalization of Eq. (\ref{eq20}) and Eq. (\ref{eq21}) as
well as their derivatives results in the equations for $C_{L}$ and the
radial coordinate of the overlap region $r_{0}$:

\begin{eqnarray}
C_{L} &=&-\frac{\sigma }{r_{0}-1}\left\{ A_{1}\sin \left[ \sigma \ln \left(
r_{0}-1\right) \right] -A_{2}\cos \left[ \sigma \ln \left( r_{0}-1\right) %
\right] \right\} ,  \notag \\
r_{0} &=&1+\exp \left[ -\sigma ^{-1}\arctan \frac{A_{1}-A_{2}\sigma }{%
A_{2}+A_{1}\sigma }\right] ,  \label{eq22}
\end{eqnarray}

\noindent where $A_{1}=2\Re \left( C_{S}\right) $, $A_{2}=-2\Im
\left( C_{S}\right) $ and can be expressed through the amplitudes
of ingoing and outgoing perturbation waves in the vicinity of the
horizon:

\begin{eqnarray}
a_{out} &=&\frac{1}{2}\left( A_{1}-iA_{2}\right) ,  \notag \\
a_{in} &=&\frac{1}{2}\left( A_{1}+iA_{2}\right) .  \label{eq23}
\end{eqnarray}

Reflection coefficient $R\equiv a_{out}\diagup a_{in}$ ($\left\vert
R\right\vert =1$, that corresponds to nonabsorbing surface, see above), $%
a_{in}$ is the amplitude of the ingoing wave transmitted through the
Regge-Wheeler potential barrier \cite{wheeler}. If we neglect the momentum
transmitted by the ingoing wave to the collapsar, it is possible to choose $%
A_{1}=0$ or $A_{2}=0$ (i.e. the phase shift due to reflection is equal to $%
\pi $).

If $A_{1}=0$, then

\noindent
\begin{eqnarray}
C_{L} &=&\frac{\sigma }{r_{0}-1}A_{2}\cos \left[ \sigma \ln \left(
r_{0}-1\right) \right] ,  \notag \\
r_{0} &=&1+\exp \left[ -\sigma ^{-1}\arctan \left( -\sigma \right) \right] .
\label{eq24}
\end{eqnarray}

\noindent If $A_{2}=0$, then

\begin{eqnarray}
C_{L} &=&-\frac{\sigma }{r_{0}-1}A_{1}\sin \left[ \sigma \ln \left(
r_{0}-1\right) \right] ,  \notag \\
r_{0} &=&1+\exp \left[ -\sigma ^{-1}\arctan \left( \sigma ^{-1}\right) %
\right] .  \label{eq25}
\end{eqnarray}

\noindent It is clear that only Eq. (\ref{eq25}) provides the overlap of the
Logunov's and the Schwarzschild's regions in the low-frequency limit.

\section{Perturbation modes}

As it was mentioned, our analysis is based on the following main
assumptions: 1) $\sigma $ is not pure imaginary (see Appendix); 2)
$\Delta \longrightarrow 0 $ and $\Delta ^{2}=O\left( \epsilon
^{2}\right) ,$ $\epsilon \Delta =O\left( \epsilon ^{2}\right) $.
Only in this case the angular and the radial variables are
separable in the vicinity of horizon. Such separability allows
finding the horizon asymptotic of the axial perturbations, when
$\epsilon \neq 0$. Such asymptotic differs radically from that in
the Schwarzschild case: collapsar horizon is
\textit{nonabsorbing}. \qquad

As a result, there exist following main types of the perturbation
modes: 1) \textit{unbounded modes}: (i) propagating through the
Regge-Wheeler potential barrier, escaping the vicinity of the
collapsar after reflection from its surface and the subsequent
backward propagation through the barrier; (ii) reflecting from the
Regge-Wheeler potential barrier; 2) \textit{bounded modes}, which
are resonantly trapped between the collapsar surface and the
Regge-Wheeler potential barrier.

The latter modes are most astonishing as the vicinity of collapsar behaves
like resonator accumulating the gravitational perturbations with the
resonant frequencies. As a consequence, the band-gaps in the frequencies of
the reflected modes have to exist.

The trapped modes can be calculated on the simple basis. In the
correspondence with the previous analysis the trapped mode has the
zero amplitude on the horizon and the zero asymptotic far off the
Regge-Wheeler potential. In the tortoise radial coordinate we have
the usual stationary Schr\"{o}dinger equation for the modes
trapped between collapsar and the Regge-Wheller potential formed
by the curvature of the effective Riemannian spacetime
\cite{wheeler}. $\sigma $ corresponding to the trapped modes are
given in Table \ref{T1} (angular momentum is defined through
$L=l-1$).

\begin{table}
  \centering
  \caption{Frequencies of the trapped perturbations with different angular momentum, $\chi =10^{-44}$.}\label{T1}
\begin{tabular}{|c|c|c|}
\hline
$L=0$ & $L=1$ & $L=2$ \\ \hline
0.03504283094 &  0.07765951326 &  0.08629600222 \\ \hline
&  0.1499033022 &  0.1588426895 \\ \hline
&  0.2218760915 &  0.2321809639 \\ \hline
&  0.2933223483 &  0.3052900260 \\ \hline
&  0.3642142227 &  0.3780251314 \\ \hline
&  0.4344882047 &  0.4503665174 \\ \hline
&  &  0.5223159963 \\ \hline
&  &  0.5938754078 \\ \hline
\end{tabular}
\end{table}

\section{Conclusion}

\noindent Approximated horizon solutions for the axial
perturbations of the spherically symmetric metric have been
obtained in the RTG framework. It was shown, that there is no
absorption of the perturbations by the horizon. The reflection of
the gravitational waves forms the modes trapped between the
horizon and the Regge-Wheeler potential barrier. Hence the
collapsar can resemble rather the gravitational \textquotedblleft
resonator\textquotedblright or \textquotedblleft
mirror\textquotedblright than the \textquotedblleft
hole\textquotedblright.

Author appreciates S.L.Cherkas (Institute of Nuclear Problems,
Minsk, Belarus) for the valuable and stimulating discussions.

This work has been realized in the Maple 9 computer algebra system
\cite{kalashnikov3}.

\section*{Appendix}

\noindent Let's be based on Eqs. (\ref{eq9},\ref{eq10}), linearize
theirs on $\epsilon$ and suppose that $\Delta=O(\epsilon)$ (or
smaller). We keep the terms of $O(\epsilon/\Delta)$ then the
resulting equation for the function $Q$ is:

\begin{eqnarray}
\frac{\partial }{{\partial r}}\left\{ {\frac{\Delta }{{r^4
}}\left[ {\frac{{\partial Q}}{{\partial r}} - \frac{\varepsilon
}{Q}\frac{\partial }{{\partial r}}\frac{{Q^2 r^2 }}{\Delta }}
\right]} \right\} + \Sigma \frac{{\sigma ^2 \sin ^3 \theta }}{{r^4
}}\frac{\partial }{{\partial \theta }}\frac{{\frac{{\partial
Q}}{{\partial \theta }}}}{{\sin ^3 \theta }} + \frac{{Q\sigma ^2
}}{\Delta } = \nonumber \\  i\sin ^3 \theta \frac{{\partial ^2
\widetilde{\omega } }} {{\partial r\partial \theta }}\sigma \left(
{1 - \frac{\Sigma } {{1 + \frac{{r^2 }} {2}\frac{\varepsilon }
{\Delta }}}} \right),
\end{eqnarray}

\noindent where \[ \Sigma  = \frac{{1 + \frac{{r^2
}}{2}\frac{\varepsilon }{\Delta }}}{{\sigma ^2  +
\frac{\varepsilon }{\Delta }\frac{{\left( { - 8r^3 \left( {r^3  -
1} \right) + 5 - 6r} \right)}}{{4r^2 }}}}.
\]

If $\widetilde{\omega }=0$ and $\epsilon/\Delta$ is the term of
order $O(1)$ or higher then Eq. (26) results in:

\begin{equation}
\left( {1 + \frac{\varepsilon }{\Delta }} \right)\frac{{\partial
Q}}{{\partial r}} + \frac{{1 + \frac{\varepsilon }{{2\Delta
}}}}{{\sigma ^2  - \frac{\varepsilon }{{4\Delta }}}}\sigma ^2 \sin
^3 \theta \frac{\partial }{{\partial \theta
}}\frac{{\frac{{\partial Q}}{{\partial \theta }}}}{{\sin ^3 \theta
}} + \left( {\frac{{\sigma ^2 }}{\Delta } - \frac{\varepsilon
}{{\Delta ^2 }}} \right)Q = 0.
\end{equation}

After separation of the variables and keeping the leading terms we
have for the radial function:

\begin{equation}
\frac{{d\Xi }}{{dr}} + \Xi \left[ {\frac{{\sigma ^2 }}{\Delta } -
n - \frac{\varepsilon }{\Delta }\left( {\frac{{1 + \sigma ^2
}}{\Delta } + \frac{n}{2}\left( {\frac{1}{{2\sigma ^2 }} - 1}
\right)} \right)} \right] = 0.
\end{equation}

Integrating this equation, neglecting $\epsilon$ and keeping
$\epsilon/\Delta$ gives:

\begin{equation}
\Xi  = C_L \left( {r - 1} \right)^{ - \sigma ^2 } r^{\sigma ^2 }
\exp \left[ {nr - \frac{\varepsilon }{\Delta }\left( {1 + \sigma
^2 } \right)} \right],
\end{equation}

\noindent that demonstrates the strong near-horizon suppression of
the perturbation amplitude. In the vicinity of some point, which
is displaced from the horizon on $\alpha =o(\epsilon)$, we have
the leading term:

\begin{equation}
\Xi  = C_L \left( {r - \left( {1 + \alpha } \right)} \right).
\end{equation}

More rigorous analysis of the axial perturbations needs a
numerical approach.

Stability of the metric under consideration needs some additional
comments. Eqs. (19) and (29) admit the purely imaginary $\sigma$.
However in the Schwarzschild's case, the time-growing ingoing
near-horizon solution ($(r - 1)^{ - i\sigma }  = (r - 1)^\gamma
\xrightarrow[{r \to 1}]{}0$ for $\gamma>0$ providing $\exp \left(
{\gamma t} \right)\xrightarrow[{t \to \infty }]{}\infty $) can not
be matched to the time-damped outgoing at infinity solution
\cite{vish}. The RTG near-horizon solution has to be matched to
the GR near-horizon solution. As the last doesn't exist for the
case of the ingoing perturbation with $\sigma=i\gamma$ and
$\gamma>0$, the metric under consideration is stable against such
perturbation. However, in contrast to the GR, there exist the
near-horizon outgoing perturbations, which, in principle, allow
matching with the Schwarzschild's asymptotic at infinity even for
$\sigma=i\gamma$ ($\gamma>0$). When $\widetilde{\omega }= 0$, such
perturbations don't permit the matching between the GR and the RTG
solutions as long as $\gamma\leq 1$ (see Eq. (29)). Case of
$\gamma > 1$ doesn't provide the finite asymptotic on horizon.
Thus, the stability of the metric against outgoing perturbation
can be accepted as proved for $\widetilde{\omega }= 0$.

The case of $\widetilde{\omega }\neq 0$ needs an additional
consideration, which is difficult in the RTG because it is not
possible to reduce the problem to an analysis of the single
perturbation function (see Eq. (26)).


\bigskip

\begin{thebibliography}{99}
\bibitem{logunov1} A.A.Logunov, \textquotedblleft Relativistic theory of
gravity\textquotedblright , Nova Sc. Publ., 1998.

\bibitem{logunov2} A.A.Logunov, M.A.Mestvirishvili, \textquotedblleft
Relativistic theory of gravitation\textquotedblright , Nauka, Moskow, 1989
(in russian).

\bibitem{logunov3} A.A.Logunov, \textquotedblleft The theory of
gravitational field\textquotedblright , Nauka, Moskow, 2000 (in russian).

\bibitem{logunov4} A.A.Logunov, \textquotedblleft The theory of
gravity\textquotedblright , arXiv:gr-qc/0210005.

\bibitem{logunov5} S.S.Gerstein, A.A.Logunov, M.M.Mestvirishvili,
\textquotedblleft Graviton mass and total relative density of mass $\Omega
_{tot}$ in Universe\textquotedblright , arXiv:astro-ph/0302412.

\bibitem{kalashnikov1} V.L.Kalashnikov, \textquotedblleft Quintessential
cosmological scenarious in the relativistic theory of
gravitation\textquotedblright , arXiv:gr-qc/0208070.

\bibitem{talmadge} C.Talmadge, et al. Phys. Rev. Lett.
\textbf{61}, 1159 (1988).

\bibitem{logunov6} A.A.Logunov, M.A.Mestvirishvili, \textquotedblleft What
happens in the vicinity of the Schwarzschild sphere when nonzero gravitation
rest mass is present\textquotedblright , arXiv:gr-qc/9907021.

\bibitem{kalashnikov2} V.L.Kalashnikov, \textquotedblleft Static spherically
symmetric collapsar in the relativistic theory of gravitation: numerical
approach and stability analysis\textquotedblright , astro-ph/0304466.

\bibitem{fock} V.Fock, \textquotedblleft Three lectures on relativity
theory\textquotedblright , Rev. Mod. Phys. \textbf{29}, 325 (1957).

\bibitem{chandra} S.Chandrasekhar, \textquotedblleft The mathematical theory
of black holes\textquotedblright , Clarendon Press, 1983.

\bibitem{wheeler} T.Regge, J.A.Wheeler, \textquotedblleft Stability of
Schwarzschild singularity\textquotedblright , Phys. Rev. \textbf{108}, 1063
(1957).

\bibitem{sanchez} N.S\'{a}nchez, \textquotedblleft Wave scattering theory
and the absorption problem for a black hole\textquotedblright , Phys. Rev. D
\textbf{16}, 937 (1977).

\bibitem{kalashnikov3}http:$\backslash \backslash$ares.photonik.tuwien.ac.at$\backslash$BHperturbations.html

\bibitem{vish} C.V.Vishveshwara,\textquotedblleft Stability of the
Schwarzschild metric\textquotedblright , Phys. Rev. D \textbf{1},
2870 (1970).

\end{thebibliography}
\end{document}